\documentclass{emulateapj}
\shorttitle{FRBs from NS-NS inspiral} \shortauthors{Wang, Yang, Wu,
Dai \& Wang}

\begin{document}

\title{Fast Radio Bursts from the Inspiral of Double Neutron Stars}

\author{Jie-Shuang Wang$^{1,2}$, Yuan-Pei Yang$^{1,2}$, Xue-Feng Wu$^{3,4}$, Zi-Gao Dai$^{1,2}$, Fa-Yin Wang$^{1,2}$}

\affil{$^1$School of Astronomy and Space Science, Nanjing University, Nanjing 210093, China; dzg@nju.edu.cn\\
$^2$Key Laboratory of Modern Astronomy and Astrophysics (Nanjing University), Ministry of Education, China\\
$^3$Purple Mountain Observatory, Chinese Academy of Sciences, Nanjing 210008, China\\
$^4$Joint Center for Particle, Nuclear Physics and Cosmology, Nanjing
University-Purple Mountain Observatory, Nanjing 210008, China}

\newcommand{\be}{\begin{equation}}
\newcommand{\ee}{\end{equation}}
\newcommand{\g}{\gamma}
\def\ba{\begin{eqnarray}}
\def\ea{\end{eqnarray}}
\def\cE{{\cal E}}
\def\cR{{\cal R}}
\def\bmu{{\mbox{\boldmath $\mu$}}}
\def\bphi{{\mbox{\boldmath $\phi$}}}

\begin{abstract}
In this paper we propose that a fast radio burst (FRB) could originate from
the magnetic interaction between double neutron stars (NSs) during their final inspiral
within the framework of a unipolar inductor model. In this model,
an electromotive force is induced on one NS to accelerate electrons to an
ultra-relativistic speed instantaneously. We show that coherent curvature radiation
from these electrons moving along magnetic field lines in the magnetosphere of the other NS
is responsible for the observed FRB signal, that is, the characteristic emission frequency,
luminosity, duration and event rate of FRBs can be well understood. In addition, we
discuss several implications of this model, including double-peaked FRBs and possible
associations of FRBs with short-duration gamma-ray bursts and gravitational wave events.
\end{abstract}


\section{Introduction}
Fast radio bursts (FRBs) are flashes with durations of order 1\,ms at
typical frequencies of $\sim 1$\,GHz. Up to now, 17 FRBs have been detected
\citep{Lorimer2007,Keane2012,Thornton2013,Burke2014,Spitler2014,Spitler2016,Ravi2015,Petroff2015,Champion2015,Masui2015,Keane2016}.
Recently, a new burst (FRB 150418) and its fading radio transient
lasting $\sim6$ days were reported \citep{Keane2016}. The host
galaxy associated with this transient has been identified to be an
elliptical galaxy with redshift $z=0.492\pm0.008$.

However, whether or not this radio transient is an afterglow of FRB 150418 remains controversial. On one hand,
\cite{Williams2016} and \cite{Vedantham2016} argued that the radio transient may be faring
from an active galactic nucleus. On the other hand, \cite{LiZhang2016} statistically examined the chance
coincidence probability to produce such a transient and found that the possibility of being an afterglow of
FRB 150418 is not ruled out. In despite of this debate, the observed dispersion measures, which are
in the range of a few hundreds to few thousands pc\,cm$^{-3}$, still provide strong evidence that
at least some FRBs including FRB 150418 are at cosmological distances.

Many models have been proposed to explain the properties of FRBs, including
giant flares from magnetars \citep{Popov2010,Kulkarni2014}, annihilations of
mini-black holes (BHs) \citep{Keane2012}, mergers of two neutron stars (NSs)
\citep{Totani2013}, NS-BH mergers \citep{Mingarelli2015},
double white dwarf mergers \citep{Kashiyama2013}, blitzars
\citep{Falcke2014,Zhang2014}, eruptions of nearby flaring stars \citep{Loeb2014},
collisions between NSs and asteroids/comets \citep{Geng2015}, giant pulses from
pulsars \citep{Connor2016,Cordes2016}, and charged BH-BH mergers \citep{Zhang2016a}.
However, as argued by \cite{Zhang2016}, some of these models are clearly inconsistent with a high kinetic energy
required by the radio transient after FRB 150418 if this afterglow is indeed true.

Although observations of FRB 110523 associated with a dense magnetized plasma \citep{Masui2015}
and the repeating FRB 121102 \citep{Spitler2014,Spitler2016,Scholz2016}
seem to disfavor catastrophic event models including old NSs, as indicated by classification
of gamma-ray bursts (GRBs), at least two distinct classes of FRBs should exist \citep{Keane2016}. This conclusion is
also suggested by a recent statistic analysis \citep{Li2016}. Furthermore, the plausible radio afterglow
of FRB 150418 associated with an elliptical galaxy shows that this FRB is likely to be contemporaneous
with a short GRB, and thus the NS-NS merger model of FRBs is still favored \citep{Keane2016,Zhang2016}.

In the NS-NS merger scenario, the physical mechanism of FRBs remains a mystery.
In this {\em Letter}, we study the physical processes of forming an FRB within the
framework of a unipolar inductor model.
This model has been proposed to describe a kind of magnetic interaction in NS-NS
systems, by which energy can be extracted during the binary inspiral
\citep{Piro2012,Lai2012}. This leads to electromagnetic radiation before the
final merger. Even though \cite{Totani2013} and \cite{Mingarelli2015}
argued that the mergers of NS-NS/BH binaries could be progenitors of FRBs,
they have not carried out an analysis and modeling of relevant physical processes.
In addition, our work is different from \cite{Hansen2001}.
These authors assumed a pulsar-like coherent radio radiation during the NS-NS inspiral whereas
we here focus on the energy extraction in the unipolar inductor model.
The orientation of radio radiation is quite different between our and Hansen \& Lyutikov's
models. This in fact implies different observational associations of FRBs with other
counterparts like short GRBs.

This paper is organized as follows. In Section 2, we briefly
introduce the unipolar inductor model based on the NS-NS merger scenario. In Section 3,
we apply it to an FRB and constrain the model parameters.
In Section 4, we present our summary and discussions.

\section{The unipolar inductor model}

The unipolar inductor model was originally put forward in the Jupiter-Io system
\citep{GoldreichL1969}. This model was then applied to several systems, such as
double white dwarf binaries \citep{Wu2002} and NS-NS/BH binaries
\citep{Hansen2001,McWilliams2011,Piro2012,Lai2012}. Usually, one of two NSs in this model
has a much stronger magnetic field than that of the other NS. Recent simulations show that this model
is still likely to be established, even if the ratio of
the magnetic field strengths of two NSs is $\sim 100$ \citep{Palenzuela2013}.

\begin{figure}
\begin{center}
\includegraphics[scale=0.35]{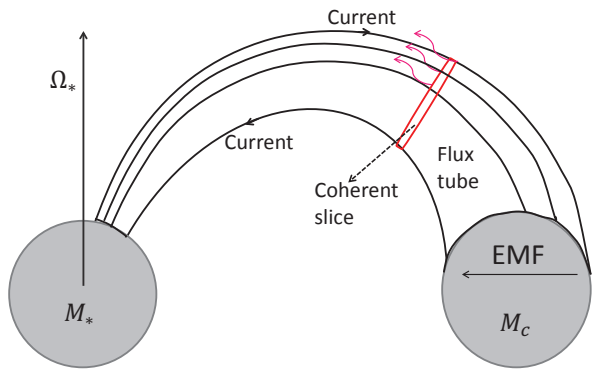}
\caption{Schematic picture of an electric circuit based on the unipolar inductor
model during the final inspiral of double neutron stars. The red block is a slice
where curvature radiation of electrons is coherent. \label{fig:fig1}}
\end{center}
\end{figure}

Considering a primary NS with mass $M_*$, spin $\Omega_*$, magnetic dipole moment
$\mu_*=B_*R_*^3$ and radius $R_*$, and its companion NS with mass $M_c$ and radius $R_c$.
The distance between the two NSs is $a$ and the orbital angular velocity is $\Omega$. We
assume that ${\bf\Omega_*}$, ${\bmu_*}$ and ${\bf\Omega}$ are aligned for simplicity. As the
companion crosses the magnetic field lines of the primary NS, an electromotive force
(EMF) is generated as shown in Fig.\,\ref{fig:fig1}, $\cE \simeq 2R_c|{\bf E}|$, where
${\bf E}= \textbf{v}\times {\bf B}/c$, ${\bf B}=\bmu_*/a^3$ and ${\bf v=(\Omega-\Omega_*)}
\times \textbf{\emph{a}}$ \citep{Piro2012,Lai2012}. During the inspiralling stage,
spin-orbit synchronization cannot be achieved by magnetic and tidal torques
\citep{Bildsten1992,Kochanek1992,Lai1994,Ho1999,Lai2012}, so the spinning angular velocity
of the primary NS could be much smaller than the orbital angular velocity and
we assume that $|{\bf v}|\sim\Omega a$. Thus the EMF is given by
\be
\cE\simeq {2\mu R_c\over ca^2}\Omega.
\label{eq:emf}\ee

The magnetic field lines exhibit like electric wires and thus a DC circuit is
established. The resistance of the circuit $(\cR_{\rm tot})$ is generally believed
to be dominated by the magnetosphere, which is in units of the impedance of free space,
namely $\cR_{\rm tot}\simeq2\cR_{\rm mag}=8\pi/c$ \citep{Piro2012,Lai2012}. As we
show below, this magnitude of resistance is reasonable.
The factor 2 is introduced because the circuit could be established on both
sides of the binary orbit. However, the circuit is unstable,
since a toroidal magnetic field is produced and grows even to be comparable
to the poloidal field. Thus, \cite{Lai2012} proposed a quasi-cyclic circuit model,
in which when the toroidal magnetic field is strong enough, the circuit
breaks down and then the magnetic energy is released due to the reconnection process.
Subsequently, the toroidal magnetic field becomes weak and the whole cycle repeats.
In this case, the energy dissipation rate is \citep{Lai2012}
\ba
\dot E_{\rm diss}= 1.7\times10^{42}B_{*,12}^2 a_{30}^{-7}\,{\rm erg/s},\label{eq:diss}
\ea
where $B_{*,12}=B_*/10^{12}$\,G and $a_{30}=a/30$\,km. This is consistent with the
energy dissipation rate given by the simulation of \cite{Palenzuela2013}. Thus
we believe that the resistance used here is reasonable.

Even if a small fraction of the binary orbital energy is extracted by the magnetic interaction,
evolution of the NS-NS binary is dominated by gravitational wave radiation
\citep{Lai2012}, following
\be
\dot a=-{64G^3M_*^3q(1+q)\over 5 c^5a^3},
\ee
where $q=M_c/M_*$ is the mass ratio of the two stars \citep{Landau1975}. For simplicity, we
set $M_*=1.4M_{\odot}$, $R_*=R_c=10$\,km and $q=1$. Thus we obtain
\be
a=a_0\left[1-{256G^3M_*^3q(1+q)\over 5 a_0^4c^5}t\right]^{1/4}=20(1-1695t)^{1/4}{\rm km},\label{a}
\ee
where we set the zero time when the surfaces of the two NSs just touch with each other,
i.e. $a=20$\,km at $t=0$. Combining with Equation\,(\ref{eq:diss}), we find the energy
dissipation rate grows drastically in the last few milliseconds, which is consistent with
the typical duration of an FRB.

\section{Modeling an FRB}

{As described by \cite{GoldreichJ1969}, the magnetosphere of a pulsar is filled
with electrons/positrons. However, due to the EMF near the surface of the companion in our model, more
electrons/positrons might be produced. We assume that the electron density $n_e$ caused by the
EMF is analogous to the space-charge density in the magnetosphere of a pulsar, and thus find
\be
n_e\simeq {{\bf\Omega}\cdot{\bf B}\over 2\pi ec}\simeq 1.5\times10^{12}B_{*,12}a_{30}^{-9/2}\,{\rm cm}^{-3},\label{eq:ne}
\ee
where we have taken the orbital angular velocity to be $\Omega=[GM_\ast(1+q)/a^3]^{1/2}$.

The EMF will accelerate these electrons/positrons, while
radiation cools them down. On the surface of the companion,
the drift velocity of the electrons is $v_d\sim cE_{\parallel}/B$, where
$E_\parallel$ is the parallel component of the electric field along the magnetic
field, which has a small value for the magnetic field configuration in our model.
Thus, the drift velocity $v_d\lesssim c$ and the curvature radius of the
electron motion could be approximately the synchrotron gyration radius. The
electron energy after acceleration could be determined by the balance between the
``synchrotron-like'' cooling and electric field acceleration, that is,
\be
P_{\rm syn}\simeq\frac{1}{6\pi}\sigma_Tc\gamma^2B^2\sim eEc,
\ee
where $\sigma_T=6.65\times10^{-25}$cm$^2$ is the Thomson scattering cross-section.
Therefore, the maximum Lorentz factor, at which the electrons are accelerated by the EMF,
is approximately given by
\be
\gamma_{\rm max}\sim\left(\frac{6\pi e\Omega a}{c\sigma_TB}\right)^{1/2}\simeq 370B_{\ast,12}^{-1/2}a_{30}^{5/4}.
\ee
The acceleration time is $t_{\rm acc}\simeq \gamma_{\rm max}m_ec/eE=1.5\times10^{-15}B_{\ast,12}^{-3/2}a_{30}^{19/4}$\,s.
This is far smaller than the orbit period, which is generally of order $\sim 1$~ms.

The characteristic frequency of curvature radiation is
\be
\nu_{\rm curv}={3c\gamma^3\over 4\pi \rho}=2.4\times10^3\gamma^3 \rho_{30}^{-1}\,\rm{Hz},
\label{eq:nu_curv}\ee
where $\rho=30\rho_{30}$\,km is the curvature radius, and $\g$ is the Lorentz factor of an emitting
electron. For a typical FRB, we have
\be
\g=75\rho_{30}^{1/3}\nu_{\rm curv,9}^{1/3},\label{eq:g}
\ee
where $\nu_{\rm curv,9}=\nu_{\rm curv}/10^9$\,Hz.
We then discuss whether or not a photon can propagate through the plasma in the magnetosphere
by considering three effects. First, the plasma frequency in the emission region is
\be
\nu_p=\frac{1}{2\pi}\left(\frac{4\pi n_e e^2}{m_e}\right)^{1/2}.\label{eq:nu_p}
\ee
In a highly-magnetized magnetosphere, the plasma effect is negligible
if $\nu_p<\g^{1/2} \nu_{\rm curv}$ \citep{Lyubarskii1998},
which is further written as
\be
a_{30}>1.0B_{*,12}^{2/9}\g_2^{-2/9}\nu_{\rm curv,9}^{-4/9},\label{eq:aup3}
\ee
where $\g_{2}=\g/10^2$ and we have taken the electron density of the emission region to be roughly
that of the acceleration region. Second, the cyclotron absorption in the magnetosphere
is generally considered in pulsar physics, whose optical depth is given by \cite{Lyubarskii1998},
\be
\tau_{\rm cyc}\simeq 2\times10^{-3}{B_{*,12}^{3/5} P_*^{-9/5}\nu_{\rm curv,9}^{-3/5}\g_2^{-3/5}},
\ee
where $P_*=2\pi/\Omega_*$ and the electron number density outside of the flux tube (see Fig. 1)
has been assumed to be described by the Goldreich-Julian density. Thus, the cyclotron absorption is ignorable
if $P_*>0.03B_{*,12}^{1/3}\nu_{\rm curv,9}^{-1/3}\g_2^{-1/3}$\,s.
Third, the characteristic distance ($l$) of a photon due to the Thomson scattering is \citep{Lyubarskii1998}
\be
l={1\over (1-\beta_e\cos\theta){\bar n}_e \sigma_T}=\frac{1.5\times10^{14}\,{\rm cm}}{{\bar n}_{e,10}(1-\beta_e\cos\theta)},
\ee
where $\beta_e c$ is the velocity of an electron, $\theta$ is the angle between the motion directions
of the electron and photon, and ${\bar n}_{e,10}={\bar n}_e/10^{10}\,{\rm cm}^{-3}$ has been assumed to be the mean electron
number density of the magnetosphere. The light cylinder $R_L$ for a pulsar with period $P_*$
is $R_L={cP_*/2\pi}=4.88\times10^9(P_*/1\,{\rm s})\,{\rm cm} \ll l$. Therefore, we conclude that a photon with frequency of order 1\,GHz
can propagate freely through the magnetosphere.

For a single relativistic electron or positron, the emission power of
curvature radiation is $P_e=2\g^4e^2c/3\rho^2$. The electron cooling timescale
due to curvature radiation is
\be
t_{\rm cool}={\g m_e c^2\over P_e}=1.6\times10^9\rho^{2}_{30}\g_{2}^{-3}\,{\rm s}.
\ee
An electron will spend a typical time $t_{\rm circuit}\sim a/c=0.1$\,ms
in moving from the companion to the primary star. Thus the electron will not be cooled down
by curvature radiation. This hints that once an electric circuit is established (or broke up),
electrons will fill (or leave) the flux tube (see Fig.\,\ref{fig:fig1}) immediately.
Therefore, it is reasonable to assume that the total volume of the emitting region is approximately
given by $V_{\rm tot}\simeq2V_{\rm FT}\sim\pi^2 R_c^2a/3$, where the factor 2 is due to the fact
that the flux tube is established on both sides of the binary orbit, and the volume of the flux tube
(approximated by a circular cone with height $\pi a/2$) is taken to be $V_{\rm FT}\sim\pi^2 R_c^2a/6$.

However, once the wavelength of radiation is comparable to the size of
the emission region, the emission is coherent \citep{Ruderman1975}. In this case, the
volume of a coherent slice (see Fig.\,\ref{fig:fig1}) is roughly $V_{\rm coh}\sim \pi R^2_c\times c/\nu_{\rm curv}$.
Thus, the number of slices is $N_{\rm slices}=V_{\rm tot}/V_{\rm coh}\simeq\g^3a/4\rho$, which is similar to
the estimate of \cite{Falcke2014}. The total electron number in the emission region is roughly
$N_{\rm tot}\simeq n_eV_{\rm tot}$, where $n_e$ is the electron number density of the emission region and
is assumed to be equal to that of the acceleration region. Meanwhile, what should
be noted is that for coherent radiation, the Lorentz factors of emitting electrons should be in a certain range.
Therefore, the electrons participating in coherent radiation might be smaller, that is, $N_e\simeq\epsilon N_{\rm tot}$
with $\epsilon=0.1\epsilon_{0.1}\lesssim1$. Following \cite{Falcke2014}, the total power of coherent radiation is
\be
P_c\simeq N_{\rm slices}^{-1}N_e^2 P_e\simeq 4.6\times10^{39}\g \epsilon_{0.1}^{2} B_{*,12}^{2}\rho_{30}^{-1}a_{30}^{-8}\,{\rm erg/s},
\ee
and together with equation (\ref{eq:g}), we obtain
\be
P_c\simeq 3.5\times10^{41}\epsilon_{0.1}^{2} B_{*,12}^{2}\rho_{30}^{-2/3}a_{30}^{-8}\nu_{\rm curv,9}^{1/3}\,{\rm erg/s}.\label{eq:P_c}
\ee

We now constrain the model parameters. The total power of coherent radiation should be smaller than
the energy dissipation rate due to the unipolar inductor, $P_c\lesssim \dot E_{\rm diss}$, which is also written by
\be
a_{30}\gtrsim 0.2\epsilon_{0.1}^{2}\rho_{30}^{-2/3} \nu_{\rm curv,9}^{1/3}.\label{eq:aup1}
\ee
A typical FRB generally has a luminosity $P_c\gtrsim 10^{40}$\,erg\,s$^{-1}$, which leads to
\be
a_{30}\lesssim1.6\epsilon_{0.1}^{1/4} B_{*,12}^{1/4}\rho_{30}^{-1/12}P_{c,40}^{-1/8}\nu_{\rm curv,9}^{1/24},\label{eq:aup2}
\ee
where $P_{c,40}=P_c/10^{40}\,{\rm erg}\,{\rm s}^{-1}$.
An example of constraints on $a$ and $\epsilon$ for a successful FRB from inequalities (\ref{eq:aup3}), (\ref{eq:aup1})
and (\ref{eq:aup2}) is shown in Fig.\,{\ref{fig:fig2}}.
The right-side region of the solid line is unphysical because $P_c> \dot E_{\rm diss}$,
while the left-side region of the dashed line means that FRBs may be too faint to be detected.
When the separation $a$ becomes smaller than $\sim 28$\,km, the curvature emission might be
absorbed and further turn into X-ray and/or $\gamma$-ray emission when the electrons hit the surface of the
primary star \citep{McWilliams2011}. Thus, we find that the coherent radio emission is observable
in the range of $a\sim 60$\,km to $\sim 28$\,km for a typical primary pulsar with $B_*=10^{12}$\,G and an appropriate
fraction $\epsilon$.

\begin{figure}
\begin{center}
\vspace{5mm}
\includegraphics[scale=0.6]{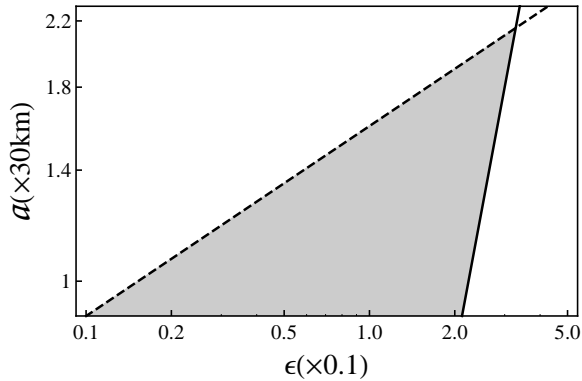}
\caption{Constraints on the parameters $a$ and $\epsilon$ with
assumptions of $B_{*,12}=1$, $\rho_{30}=1$, and $P_{c,40}=1$. The typical frequency
of observed FRBs ($\nu_{\rm curv,9}=1.4$) is taken here. Due to inequality (\ref{eq:aup3}),
the Y-axis starts from $a\simeq 28$\,km. The solid line comes from
inequality (\ref{eq:aup1}), while the dashed line comes from inequality (\ref{eq:aup2}).
\label{fig:fig2}}
\end{center}
\end{figure}

The current event rate of FRBs is roughly $\sim2.8\times10^3$\,Gpc$^{-3}$yr$^{-1}$
at redshift $z\lesssim 1$ with normalization of the daily all-sky FRB rate of order $\sim 10^4$
\citep{Zhang2016}, while the ``realistic estimate'' rate of NS-NS mergers is
$\sim10^3$\,Gpc$^{-3}$yr$^{-1}$ and the ``plausible optimistic estimate'' rate is
$\sim10^4$\,Gpc$^{-3}$yr$^{-1}$ \citep{Abadie2010}. Thus, the FRB rate is well consistent with the NS-NS merger rate.

\section{Summary and discussions}
In this {\em Letter}, we have studied the physical processes of an FRB
and explained its main features within the framework of the unipolar inductor model of
inspiralling NS-NS binaries. The companion non-magnetic NS crosses the magnetosphere of
the primary highly-magnetized NS and simultaneously produces an EMF, by which electrons
are accelerated to an ultra-relativistic speed instantaneously. These electrons then move
along magnetic field lines and generate coherent curvature radiation as shown
in Fig.\,\ref{fig:fig1}. The total power and timescale of coherent radiation are well
in agreement with a typical FRB.

Our model is clearly different from the short GRB/FRB association scenario proposed by \cite{Zhang2016a},
who suggested that the inspiral of a charged BH-BH binary forms an electric circuit and produces
an induced magnetic field. This field, if co-rotating with the BHs around the center of
mass of the binary system, would behave like a ``giant pulsar''. The coherent curvature radiation
from the magnetosphere of this pulsar-like object could explain the properties of FRBs.

Inequalities (\ref{eq:aup3}), (\ref{eq:aup1}) and (\ref{eq:aup2}) have given constraints on
the model parameters. For a larger orbital
separation or a lower magnetic field, the flux of an FRB might be too low to be observed.
For a higher magnetic field, we might observe an FRB in a different orbit period. This indicates
a double-peaked FRB if the emission during two orbit periods is observed. In addition,
there are some connections among FRBs, short GRBs/afterglows, and gravitational wave events
in our model. In what follows, we would like to discuss several implications of this model in some detail.

First, since the direction of curvature radiation might not be
aligned with the orbital angular momentum that is just the direction of a subsequent short GRB,
it is not necessary to observe an FRB and a short GRB contemporaneously. However,
it seems easier to detect an FRB associated with a short GRB afterglow, since
an afterglow generally has a wider opening angle than a GRB itself does \citep{Keane2016}.

Second, the coherent curvature radiation could not point to us at all of the times
during one orbital period. Thus, for binaries with stronger magnetic interaction,
we might observe an FRB signal during two or more periods, which might be responsible for
an FRB with double peaks \citep{Champion2015}.

Third, X-ray emission is predicted in the unipolar inductor model
\citep{Hansen2001,McWilliams2011,Lai2012,Palenzuela2013}. Following \cite{McWilliams2011},
a fraction of energy will be kept in the form of plasma kinetic energy. When this plasma
reaches the primary NS, a hot spot might be induced. However, this hot spot would be
observed only for a source at distance $\lesssim100$\,Mpc \citep{McWilliams2011}.

Fourth, the merger of double NSs after an FRB would leave behind a rapidly rotating BH
\citep{Paczynski1986,Eichler1989} or a millisecond pulsar/magnetar \citep{Dai2006,Zhang2013}.
Such objects have been widely argued as both the central engine of short GRBs and one of
the gravitational wave events that could be detected by the advanced LIGO. What should be noted
is that these two kinds of central objects have different properties of gravitational
waves, which, if detected, could further identify post-merger compact stars. Therefore,
it would be expected to see possible associations of FRBs with short GRB afterglows and
gravitational wave events in the future. The detections/non-detections of such associations
would confirm/constrain/exclude our model.

Finally, \cite{Spitler2016} and \cite{Scholz2016} recently reported their detections of ten and six
additional bright bursts from the direction of FRB 121102, respectively. This repeating FRB is obviously
distinct from the other non-repeating FRBs and thus challenges all the catastrophic event models. Very recently
we proposed a novel DC circuit model, in which a repeating FRB could originate from a highly magnetized pulsar
encountering with an asteroid belt of another star \citep{Dai2016}. During each pulsar-asteroid impact,
an electric field induced on the elongated asteroid near the pulsar's surface can drive an
electric circuit. We showed that this model can well account for all the properties of FRB 121102,
including the emission frequency, luminosity, duration, and repetitive rate. We also predicted
the occurrence rate of similar repeating sources.

\acknowledgements We thank an anonymous referee for useful comments and
constructive suggestions that have allowed us to improve our manuscript significantly.
We also thank Yong-Feng Huang, Xiang-Yu Wang and Bing Zhang for
useful discussions. This work was supported by the National Basic
Research Program (``973" Program) of China (grant No. 2014CB845800)
and the National Natural Science Foundation of China (grant Nos.
11573014, 11322328, 11433009, 11422325 and 11373022).

\end{document}